\documentclass[conference]{IEEEtran}
\IEEEoverridecommandlockouts
\usepackage{algorithm}
\usepackage{hyperref} % For hyperlinks
\usepackage{url}
\usepackage{float}

\hypersetup{
    colorlinks=true,
    linkcolor=blue,     % Color for internal links (e.g., Table of Contents)
    citecolor=blue,     % Color for citations
    urlcolor=blue       % Color for URLs
}
\usepackage{cite}
\usepackage{amsmath,amssymb,amsfonts}
\usepackage{graphicx}

\usepackage{algorithm}     % KEEP
\usepackage{algorithmic}   % KEEP

\usepackage{textcomp}
\usepackage{xcolor}
\usepackage{array}  % Add this line for better table support
\def\BibTeX{{\rm B\kern-.05em{\sc i\kern-.025em b}\kern-.08em
    T\kern-.1667em\lower.7ex\hbox{E}\kern-.125emX}}  

\IEEEspecialpapernotice{2025 IEEE 18th International Conference on Computer Science and Network Technology (CSNT)}

\begin{document}

\title{Optimized Approaches to Malware Detection: A Study of Machine Learning and Deep Learning Techniques}

\author{
    \begin{tabular}{cc}
        \begin{tabular}[t]{@{}c@{}}
            Abrar Fahim \\ 
            Department of Electrical and Electronic Engineering \\ 
            Islamic University of Technology \\ 
            Dhaka, Bangladesh \\ 
            \texttt{abrarfahim8@iut-dhaka.edu}
        \end{tabular} &
        \begin{tabular}[t]{@{}c@{}}
            Shamik Dey \\ 
            Department of Electrical and Electronic Engineering \\ 
            Ahsanullah University of Science and Technology \\ 
            Dhaka, Bangladesh \\ 
            \texttt{shamikdey7@gmail.com}
        \end{tabular} \\
        \\
        \begin{tabular}[t]{@{}c@{}}
            Md. Nurul Absur \\ 
            Department of Computer Science \\ 
            City University of New York \\ 
            New York, USA \\ 
            \texttt{mabsur@gradcenter.cuny.edu}
        \end{tabular} &
        \begin{tabular}[t]{@{}c@{}}
            Md Kamrul Siam \\ 
            Department of Computer Science \\ 
            New York Institute of Technology \\ 
            New York, New York, USA \\
            \texttt{ksiam01@nyit.edu} \\ 
        \end{tabular} \\
        \\
        \begin{tabular}[t]{@{}c@{}}
            Md. Tahmidul Huque \\ 
            Department of Computer Science and Engineering \\ 
            Bangladesh University of Business and Technology \\ 
            Dhaka, Bangladesh \\ 
            \texttt{md.tahmidulhuque@gmail.com}
        \end{tabular} &
        \begin{tabular}[t]{@{}c@{}}
            Jafreen Jafor Godhuli \\ 
            Department of Computer Science and Engineering \\ 
            Bangladesh University of Business and Technology \\ 
            Dhaka, Bangladesh \\ 
            \texttt{jafreenjaforgodhuli@gmail.com}
        \end{tabular} \\
    \end{tabular}
}
\maketitle

\begin{center}
    \footnotesize
    978-8-3315-3193-5/25/\$31.00~\copyright~2025 IEEE\\
    DOI: \href{https://doi.org/10.1109/CSNT64827.2025.10968061}{10.1109/CSNT64827.2025.10968061}
\end{center}

\begin{abstract}
Digital systems find it challenging to keep up with cybersecurity threats. The daily emergence of more than 560,000 new malware strains poses significant hazards to the digital ecosystem. The traditional malware detection methods fail to operate properly and yield high false positive rates with low accuracy of the protection system. This study explores the ways in which malware can be detected using these machine learning (ML) and deep learning (DL) approaches to address those shortcomings. This study also includes a systematic comparison of the performance of some of the widely used ML models, such as random forest, multi-layer perceptron (MLP), and deep neural network (DNN), for determining the effectiveness of the domain of modern malware threat systems. We use a considerable-sized database from Kaggle, which has undergone optimized feature selection and preprocessing to improve model performance. Our finding suggests that the DNN model outperformed the other traditional models with the highest training accuracy of 99.92\% and an almost perfect AUC score. Furthermore, the feature selection and preprocessing can help improve the capabilities of detection. This research makes an important contribution by analyzing the performance of the model on the performance metrics and providing insight into the effectiveness of the advanced detection techniques to build more robust and more reliable cybersecurity solutions against the growing malware threats.

% Digital systems find it hard to keep up with cybersecurity threats. The daily emergence of more than 560,000 new malware strains poses significant hazards to digital ecosystem. The traditional malware detection methods fail to operate properly and yield high false positive rates with low accuracy of the protection system. In order to address these shortcomings, this study explores ways in which malware can be detected using these machine learning (ML) and deep learning (DL). Then we systematically compare the performance of some of the widely used ML models, such as random forest, Multi-Layer Perceptron (MLP), and Deep Neural Network (DNN), for determining the effectiveness of the domain of modern malware threat systems. We use a considerable database from Kaggle, which we thoroughly preprocess and have feature-selected to increase model performance. We show that the DNN model is indeed the best traditional method because it achieved the highest accuracy of 99.92\% and an almost perfect AUC score. Furthermore, the feature selection and preprocessing can help improve the capabilities of detection. This research makes an important contribution by analyzing the performance of the model on the performance metrics and providing insight into the effectiveness of the advanced detection techniques to build more robust and more reliable cybersecurity solutions against the growing malware threats.

\end{abstract}

\section{Introduction}
\begin{figure*}
    \centering
    \includegraphics[width=\linewidth]{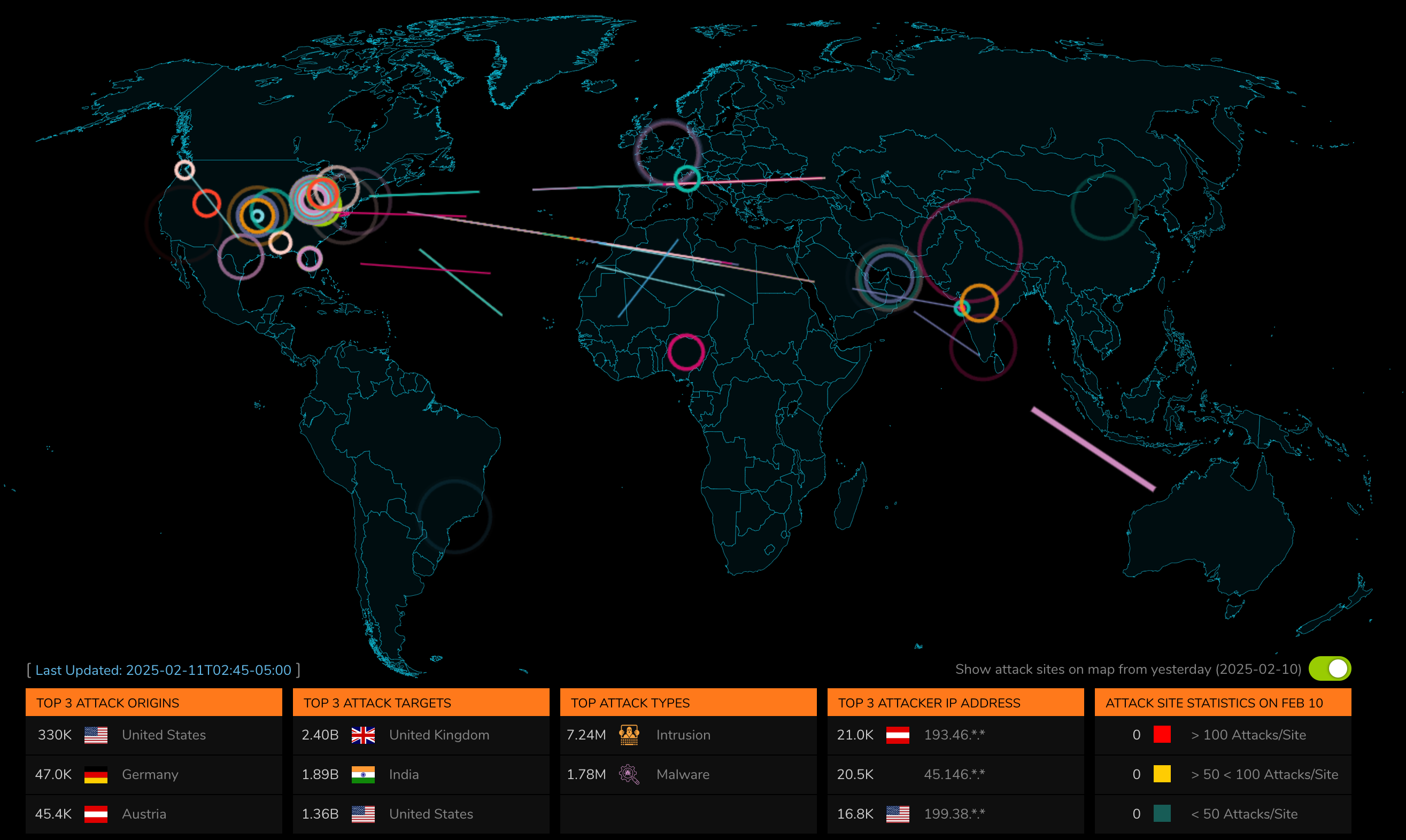}
    \caption{Attacks throughout the world~\cite{SonicWall2025}.}
    \label{fig:attacks}
\end{figure*}

According to Forbes and cybersecurity industry leaders, daily new malware variants were over 56,000 and this is enhancing the threat level to the world’s cybersecurity systems \cite{craft2024malware, Brooks2023, Astra2025}. Nevertheless, the dynamics of these threats continue to increase, whereas traditional approaches to malware identification can provide neither sufficient accuracy nor flexibility \cite{10541049,Kianpour2024 , Haque2024DDoS}. The increased vulnerability of sensitive data due to viruses, cyberattacks, cyberthreats, and cybercrimes must be addressed. The countries that encountered various attack types during a random week in 2025 are shown in Figure~\ref{fig:attacks}, along with the main malware suppliers.

Most current solutions have relatively low accuracy connected with a high number of false positives and do not allow efficient work with new, complex forms of malware. This creates a significant gap in the possibility of protection through detection since the current systems fail to recognize the modern threats that endanger the data and privacy of the people.However, the current level of ML and DL technologies has shown that malware detection has to be improved in order to fill this research gap. Due to the hardware improvements, the current ML algorithms can easily process large volumes of data and are suitable to detect any type of malware; however, they have their shortcomings, especially when implementing high-dimensional data sets or new strains of malware. In terms of features, DL models are capable of capturing and performing subtle details; hence, they can be applied to discover more enhanced malware. Moreover, there has not been sufficient effort made to compare the effectiveness of these techniques, which creates the need for a systematic approach.This research seeks to fill this gap by making the following two comparisons: (a) an assessment of the performance of different ML models for malware detection and, (b) an assessment of the performance of different DL models for malware detection. In this study, various algorithms, including Random Forest, MLP, and DNN, are evaluated to determine the best ways of enhancing the accuracy of malware detection with a focus on the modern threats.

\section{Related Works}
There has been much research conducted and is being conducted in the area of malware detection through the use of machine learning and deep learning. Also, depending on the type of malware, researchers have also explored the performance of various models and datasets for malware detection, which produced various results. The following summaries highlight key findings from notable works in this area:\\\\
A recent study used a dataset from the Canadian Institute for Cybersecurity with 17,394 instances to assess the efficacy of three disparate algorithms: CNN, DT, and SVM with respect to the identification of malware~\cite{sym14112304}. Out of all the tested models, the Decision Tree model had the highest accuracy of 99\%, whereby it clearly exhibits the capability of managing the dataset's feature space. The CNN was also impressive, achieving 98.76\% accuracy; this could be attributed to the network’s capability to identify tiered structures from data. Although its performance was not as high as the CNN with an accuracy of 96.41\%, it was still feasible for use in malware classification.\\In another study, the authors used a large number of machine learning methods to identify the Windows malware sample by using Kaggle data~\cite{10.1007/978-981-16-7618-5_53}. Among all the tested models, Random Forest was the best one, using which an overall accuracy of the model was found to be 99.44\%. As for other models, SVM Decision Tree, ADA Boost, and all these models were also considered but did not appear to work well. The study also showed that Random Forest has other strengths, including the ensemble of decision trees to minimize overfitting.\\DIVAKARLA et al. combined a new deep learning technique for Windows malware detection employing the EMBER dataset~\cite{DIVAKARLA2022148}. Their approach indeed got an accuracy of 87.76\%, which shows that their approach is promising in handling polymorphic malware. A form of malware that morphs its code to avoid being detected is quite a nightmare to current detection techniques. The authors indicate that the application of deep learning methods may yield higher performance rates while analyzing intricate manifestations of malware activity; however, improvement is still possible.\\Stacked ensemble learning was utilized by a group of researchers to conduct malware classification from the Portable Executable (PE) malware dataset of 19,611 samples~\cite{informatics8010010}. This approach outperformed by combining predictions from multiple models. The accuracy of the Random Forest model was at the highest of 99.24\%, closely followed by the Decision Tree model, which got 98.29\%. This emphasizes the possibility of ensemble learning techniques used to enhance the strength of different classifiers in a composite detection process.\\Machine learning models such as Random Forest and Gaussian Naive Bayes were explored by the author for the task of dynamic malware detection~\cite{Akhtar2023}. All training on wide PE files was taken from Kaggle. To the extent these results could be reproduced, they were achieved with an accuracy of 100\% in behavior-based detection, showing that dynamic analysis, observing malware behavior during execution, is essential in overcoming the shortcomings of static features.\\Another remarkable study presented a Hybrid Feature Malware Detection System (HF-MDS), which is built from the LSVC model by fusing different types of analysis, such as static and dynamic analysis~\cite{article}. The real-world malware precision achieved with this hybrid approach was 99.743\%, showing that integrating multiple feature sets improves detection accuracy. It was found in the study that combining static and dynamic features can capture a wider range of malware characteristics with greater performance. Strategies like as ML has a track record of positively influencing society and improving it\cite{Siam2024}.
\\Furthermore, in another work, hardware performance counters (HPCs) were used to analyze malware detection, where MLP, CNN, and Full Order Radial Basis Function (RBF) were applied\cite{bawazeer2021malware}. These models were able to report performance numbers as 96.95\%, 98.2\%, and 98.68\%. HPC-based detection methods using low-level hardware data to detect malicious activity were highlighted as potential. By comparison, the performance of the CNN model demonstrates that it has the capacity to handle the spatial patterns behind HPC data.

\section{METHODOLOGY}\label{AA}
\subsection{Data Collection}

This study utilized the Malware Detection dataset, which was acquired from Kaggle, a well-known platform for data science and machine learning contests~\cite{saravana2023malware}. This dataset focuses on PC malware detection and comprises of 100,000 data points, each characterized by 35 distinct features. The dataset includes instances classified into two categories: 'benign' and 'malware'. The features capture various attributes relevant to network traffic and malware behavior.
\subsection{Preprocessing}
Preprocessing has been defined as a critical component in data analysis and machine learning because it conditions or prepares the data so that we get the data in a proper format to influence the performance of the model\cite{AmatoDiLecce2023}. 

\begin{itemize}
    \item \textbf{Object Column Encoding}: Two object-type columns were encoded using \textit{Label Encoding}, converting categorical values into numerical format for model compatibility.
    \item \textbf{Hash Column Unhashing}: A hashed column was unhashed using the \texttt{hashlib} library to retrieve the original data.
    \item \textbf{Outlier Removal Using Z-Score}: Outliers, defined as data points with Z-scores greater than 3 or less than -3, were removed to ensure the model was not affected by extreme values.
    \item \textbf{Feature Selection (Top 25 Features)}: \textit{Recursive Feature Elimination (RFE)} was used to select the top 25 features for model training. Below table \ref{tab:1} shows the list of selected features.

% \begin{table}
% \renewcommand{\arraystretch}{1.3}
% \caption{Top 25 Selected Features}
% \label{tab:1}
% \centering
% \begin{tabular}{|c|c|}
% \hline
% \bfseries Feature Name & \bfseries Feature Name \\
% \hline
% millisecond & total\_vm \\
% state & shared\_vm \\
% usage\_counter & exec\_vm \\
% prio & reserved\_vm \\
% static\_prio & nr\_ptes \\
% normal\_prio & end\_data \\
% policy & last\_interval \\
% vm\_pgoff & nvcsw \\
% vm\_truncate\_count & nivcsw \\
% task\_size & min\_flt \\
% cached\_hole\_size & maj\_flt \\
% free\_area\_cache & fs\_excl\_counter \\
% mm\_users & \\
% \hline
% \end{tabular}
% \end{table}

\begin{table}[h!]
    \centering
    \caption{Top 25 Selected Features}
    \label{tab:1}
    \renewcommand{\arraystretch}{1.2} % Adjust row height for better readability
    \setlength{\tabcolsep}{4pt} % Adjust column spacing
    \small % Reduce font size for compact display

    \begin{tabular}{|p{4cm}|p{4cm}|} % Define custom column widths
        \hline
        \multicolumn{2}{|c|}{\textbf{Features}} \\
        \hline
        millisecond & total\_vm \\ 
        state & shared\_vm \\ 
        usage\_counter & exec\_vm \\ 
        prio & reserved\_vm \\ 
        static\_prio & nr\_ptes \\ 
        normal\_prio & end\_data \\ 
        policy & last\_interval \\ 
        vm\_pgoff & nvcsw \\ 
        vm\_truncate\_count & nivcsw \\ 
        task\_size & min\_flt \\ 
        cached\_hole\_size & maj\_flt \\ 
        free\_area\_cache & fs\_excl\_counter \\ 
        mm\_users &  \\ % Empty cell for consistency
        \hline
    \end{tabular}
\end{table}

% \begin{table}[h!]
% \centering
% \caption{Top 25 Selected Features}
% \label{tab:1}
% \renewcommand{\arraystretch}{0.8} % Adjusts row height
% \setlength{\tabcolsep}{11pt} % Adjusts column spacing for increased width
% \begin{tabular}{|p{8cm}|} % Increases column width
% \hline
% \multicolumn{1}{|c|}{\Large \textbf{Selected Features}} \\ 
% \hline
% \Large millisecond, state, usage\_counter, \\
% \Large prio, static\_prio, normal\_prio, \\
% \Large policy, vm\_pgoff, vm\_truncate\_count, \\
% \Large task\_size, cached\_hole\_size, free\_area\_cache, \\
% \Large mm\_users, map\_count, hiwater\_rss, \\
% \Large total\_vm, shared\_vm, exec\_vm, \\
% \Large reserved\_vm, nr\_ptes, end\_data, \\
% \Large last\_interval, nvcsw, nivcsw, \\
% \Large min\_flt, maj\_flt, fs\_excl\_counter, \\
% \Large lock, utime, stime, \\
% \Large gtime, cgtime, signal\_nvcsw, \\
% \Large hash\_numeric \\
% \hline
% \end{tabular}
% \end{table}

    \item \textbf{Train-Test Split (80-20)}: The dataset was split into training and testing sets, with 80\% of the data (80,000 records) used for training and 20\% (20,000 records) reserved for testing.
\end{itemize}

\begin{figure}[h!]
\centering
\includegraphics[width=1\columnwidth, height=0.17\textheight]{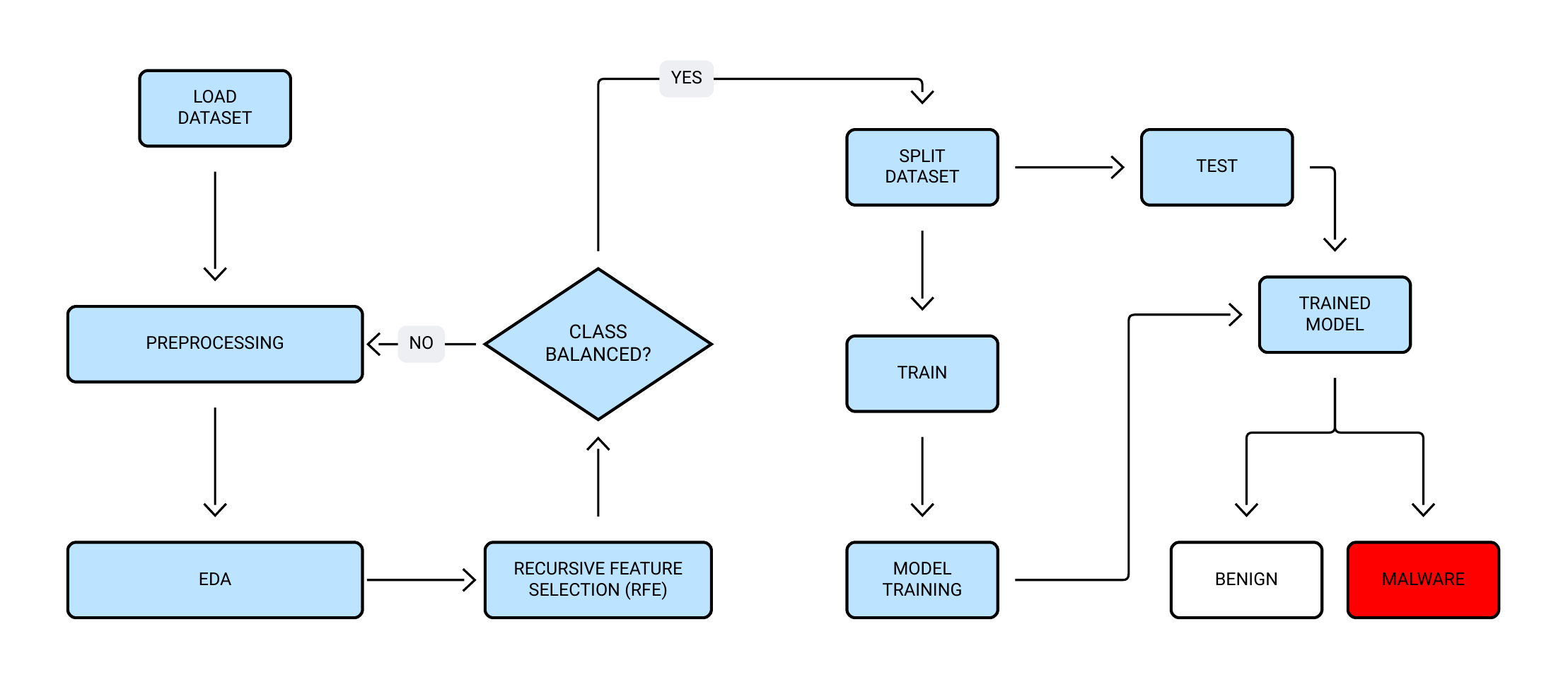}  
\caption{: Workflow Diagram }
\label{fig:methodology}
\end{figure}
\
Figure~\ref{fig:methodology} shows the workflow of the experiment, that how malware detection is processed from beginning to end.

\section{Results and Discussion}
To evaluate the performance of our models in predicting malware, we used accuracy, precision, recall, F1 score, AUC and confusion matrix as key metrics. The measure of accuracy reflects the total correctness of the model while the precision and recall measures the model’s ability to find the real positive and the degree of reliability of the positive prediction respectively. This way we have a good quality measurement and the F1 score balances precision and recall. The AUC curve measures some of the discrimination capability of the model between classes, with higher values meaning better discrimination. The confusion matrix also shows a detailed breakdown of true and false classifications as well. The results illustrate the ability of our method in malware detection, together with the merit of feature selection and the proper preprocessing in upping the model performance.

In this section we present the result of many machine learning and deep learning models in detecting malware assessment based on classification metrics.
\subsection{Machine Learning Approach}

The performance and training times of several machine learning models for malware detection are given in table \ref{tab:ml_model_accuracies}. with a highest final training accuracy of 99.77\%, 10-fold CV accuracy of 99.67\%, and a training time of 13.5 seconds the MLPClassifier was the best. With a final training accuracy of 99.97\% and on a 10 fold CV accuracy of 98.45\% it took 15.3 seconds for the KNeighborsClassifier to train. Logistic Regression had poor accuracy 93.75\%; however, it was the fastest with 12.12 seconds of training. For 10-fold CV, with a final training accuracy of 99.72\%, 99.64\% for 10-fold CV, it was very good while requiring training time of 19.6s. Final training accuracy of Random Forest was 99.60\%, CV accuracy of 99.52\% with training times of 16.3 seconds. Training times for all models were in the range of 12 to 19.6 seconds and with high accuracy.

Summary of the testing accuracies and AUC-ROC values of the machine learning models is presented in the following table \ref{tab:model_performance} for clearer classification performance of these models.

% \begin{table}[ht]
%     \centering
%     \caption{Machine Learning Model Accuracies and Training Time}
%     \begin{tabular}{|l|l|l|l|}
%         \hline
%         \textbf{Model}                     & \textbf{Training Accuracy} & \textbf{CV Accuracy(\%)} & \textbf{Training Time (seconds)} \\ 
%         \hline
%         MLP                    & 99.77                          & 99.67                       & 13.5                            \\ 
%         \hline
%         KNN              & 99.97                          & 98.45                     & 15.3                           \\ 
%         \hline
%         LR             & 93.75                          & 92.89                      & 12.12                           \\ 
%         \hline
%         SVC                                & 99.72                    & 99.64                       & 19.6                          \\ 
%         \hline
%         RF                     & 99.60                           & 99.52                      & 16.3                            \\ 
%         \hline
%     \end{tabular}
%     \label{tab:ml_model_accuracies}
% \end{table}

\begin{table}[ht]
    \centering
    \caption{Machine Learning Model Accuracies and Training Time}
    \label{tab:ml_model_accuracies}
    \renewcommand{\arraystretch}{1.2} % Adjust row height for better readability
    \setlength{\tabcolsep}{4pt} % Reduce column spacing for compact display
    \small % Reduce font size for fitting half-column

    \begin{tabular}{|p{1.5cm}|p{1.5cm}|p{2cm}|p{2.5cm}|} % Adjust column widths for half-column
        \hline
        \textbf{Model} & \textbf{Train Accuracy (\%)} & \textbf{CV Accuracy (\%)} & \textbf{Train Time (in Sec.)} \\ 
        \hline
        MLP  & 99.77 & 99.67 & 13.5 \\ 
        \hline
        KNN  & 99.97 & 98.45 & 15.3 \\ 
        \hline
        LR   & 93.75 & 92.89 & 12.12 \\ 
        \hline
        SVC  & 99.72 & 99.64 & 19.6 \\ 
        \hline
        RF   & 99.60 & 99.52 & 16.3 \\ 
        \hline
    \end{tabular}
\end{table}

This table \ref{tab:model_performance} shows malware detection machine learning model results through test accuracy and AUC score measurements. Both Multi-Layer Perceptron (MLP) and Random Forest models demonstrated excellent performance with test accuracy reaching 99.99\%. They maintained perfect scores of 100\% for AUC. The Support Vector Classifier produced a test accuracy of 99.97\% and reached a perfect AUC rating of 100\%. The K-Nearest Neighbors model showed test accuracy of 99.97\% and perfect AUC ranking at 100\%. Logistic Regression demonstrated good results with a test accuracy of 93.75\% and AUC score of 98.57\% yet smaller than other approaches. The experiments show that all examined models work well for malware detection with their high accuracy scores and AUC values.

\begin{table}[ht]
    \centering
    \caption{Model Performance Metrics}
    \begin{tabular}{|l|c|c|}
        \hline
        \textbf{Model} & \textbf{Test Accuracy(\%)} & \textbf{AUC Score(\%)} \\ 
        \hline
        Random Forest & 99.98 & 99.8 \\ 
        \hline
        SVC          & 99.97 & 99.5 \\ 
        \hline
        Logistic Regression & 93.75 & 99 \\ 
        \hline
        K-Nearest Neighbors &99.97 & 99.3 \\ 
        \hline
        Multi-Layer Perceptron & 99.99 & 99.6 \\ 
        \hline
    \end{tabular}
    \label{tab:model_performance}
\end{table}

To validate the results of the model MCC and Kappa test is performed and the results are mentioned below in Table \ref{tab:mcc_kappa}.

The table \ref{tab:mcc_kappa} contains MCC and Kappa values for multiple models. The Random Forest SVC MLP and KNN classifiers deliver outstanding performance according to very high MCC and Kappa figures close to 100\%. Logistic Regression produces results that perform well although not as highly as other models.

\begin{table}[ht]
\centering
\caption{MCC and Kappa Values for Different Models}
\begin{tabular}{|l|c|c|}
\hline
\textbf{Model} & \textbf{MCC (\%)} & \textbf{Kappa (\%)} \\
\hline
Logistic Regression & 87.51 & 87.50 \\
\hline
Random Forest & 99.94 & 99.94 \\
\hline
Multi-Layer Perceptron (MLP) & 99.99 & 99.90 \\
\hline
K-Nearest Neighbors (KNN) & 99.94 & 99.94 \\
\hline
Support Vector Classifier (SVC) & 99.95 & 99.95 \\
\hline
\end{tabular}
\label{tab:mcc_kappa}
\end{table}

\begin{figure*}[h!]
\centering
\includegraphics[width=2\columnwidth, height=0.4\textheight]{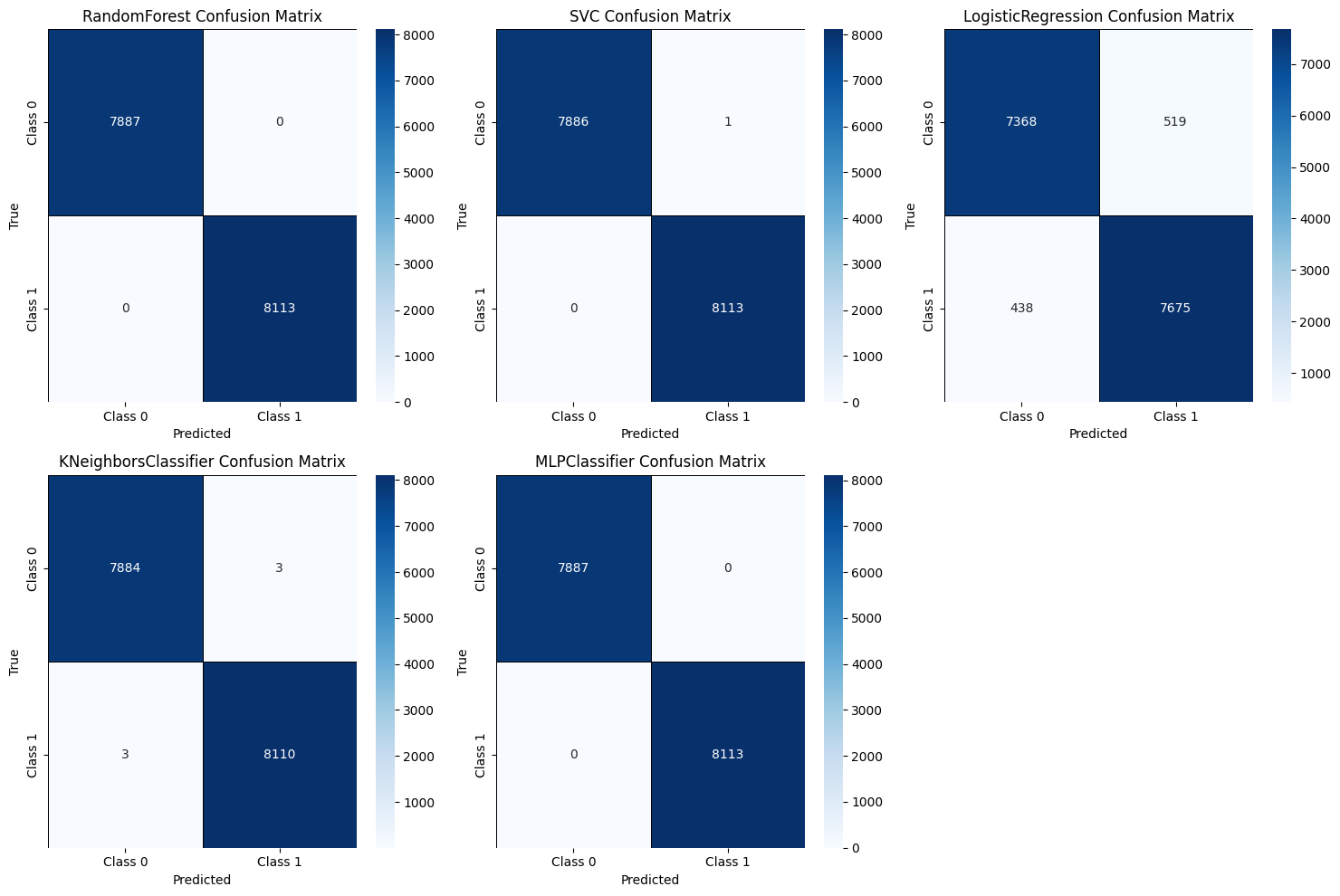}
\caption{Confusion Matrix} 
\label{fig:conf}
\end{figure*}

\begin{figure*}[h!]
\centering
\includegraphics[width=1.5\columnwidth, height=0.3\textheight]{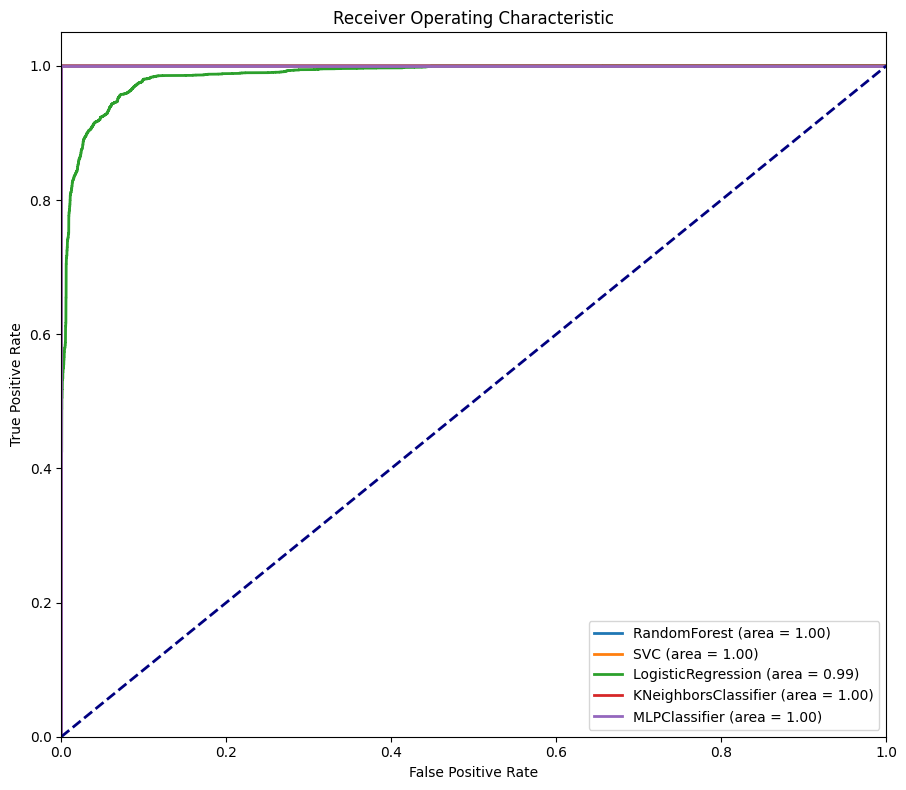}
\caption{ROC CURVE} 
\label{fig:roc}
\end{figure*}

Figure \ref{fig:conf} and \ref{fig:roc}  illustrate the confusion matrix and roc-curve of machine learning models respectively.

The figure \ref{fig:conf} suggests, the most reliable models are Random Forest (RF) and Multi-Layer Perceptron (MLP), both achieving perfect accuracy and 0 false positives.

The majority of machine learning models, as shown in Figure \ref{fig:roc}, have an AUC of 1, which denotes infallible classification.
Nevertheless, logistic regression only marginally outperforms other algorithms in class separation, with an AUC value of 0.99.

\subsection{Deep Learning Approach}
In the table \ref{tab:dl_model_accuracies}, the training performance of three deep learning models i.e., CNN, LSTM, and DNN is presented. Finally the CNN model had a high final training accuracy of 99.91\% with a 10 fold CV accuracy of 99.87\%. Training accuracy 99.90\% with CV accuracy 99.70\% was a little lower than the LSTM model. Among the rest, DNN trained with accuracy up to 99.92\% and the highest CV accuracy at 99.95\%.

\begin{table}[ht]
    \centering
    \caption{Deep Learning Model Training Performance}
    \label{tab:dl_model_accuracies}
    \renewcommand{\arraystretch}{1} % Reduces row height
    \setlength{\tabcolsep}{4pt} % Reduces column spacing
    \small % Makes text smaller

    \begin{tabular}{|p{2.8cm}|p{2.3cm}|p{2.8cm}|} % Adjust column widths
        \hline
        \textbf{Model} & \textbf{Final Training Accuracy (\%)} & \textbf{10-Fold CV Accuracy (\%)} \\
        \hline
        CNN Model  & 99.91 & 99.87 \\
        \hline
        LSTM Model & 99.90 & 99.70 \\
        \hline
        DNN Model  & 99.92 & 99.95 \\
        \hline
    \end{tabular}
\end{table}

Testing performance of CNN, LSTM and DNN has been presented in the table \ref{tab:dl_performance}. The results prove the accuracy as well as the AUC scores of each model, where the model that reaches the highest accuracy is CNN that gains 99.99\% and AUC score of 0.9889. Following an LSTM model that had an accuracy of 99.54\% and AUC of 0.9956, the accuracy achieved by the DNN model was 99.90\% and AUC of 0.9993. These results show that the performed exceptionally well with highest accuracy and perfect AUC, which means the models are effective for the current task.
\\

\begin{table}[h]
    \centering
    \caption{Deep Learning Model Testing Performance}
    \label{tab:dl_performance}
    \renewcommand{\arraystretch}{1} % Reduce row height
    \setlength{\tabcolsep}{4pt} % Reduce column spacing
    \small % Shrink text size

    \begin{tabular}{|p{2.8cm}|p{2.3cm}|p{2.3cm}|} % Match width of the previous table
        \hline
        \textbf{Model} & \textbf{Accuracy (\%)} & \textbf{AUC(\%)} \\ 
        \hline
        CNN  & 99.99 & 98.89 \\ 
        \hline
        LSTM & 99.54 & 99.56 \\ 
        \hline
        DNN  & 99.90 & 100 \\ 
        \hline
    \end{tabular}
\end{table}

\begin{figure*}[h!]
\centering
\includegraphics[width=1.5\columnwidth, height=0.3\textheight]{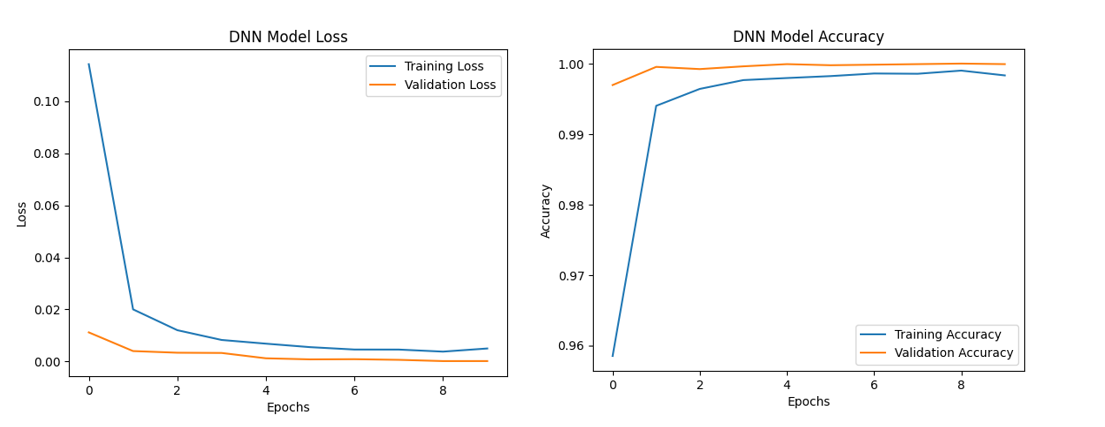}
\caption{Validation loss and Accuracy curve of DNN} 
\label{fig:val1}
\end{figure*}

\begin{figure*}[h!]
\centering
\includegraphics[width=1.5\columnwidth, height=0.3\textheight]{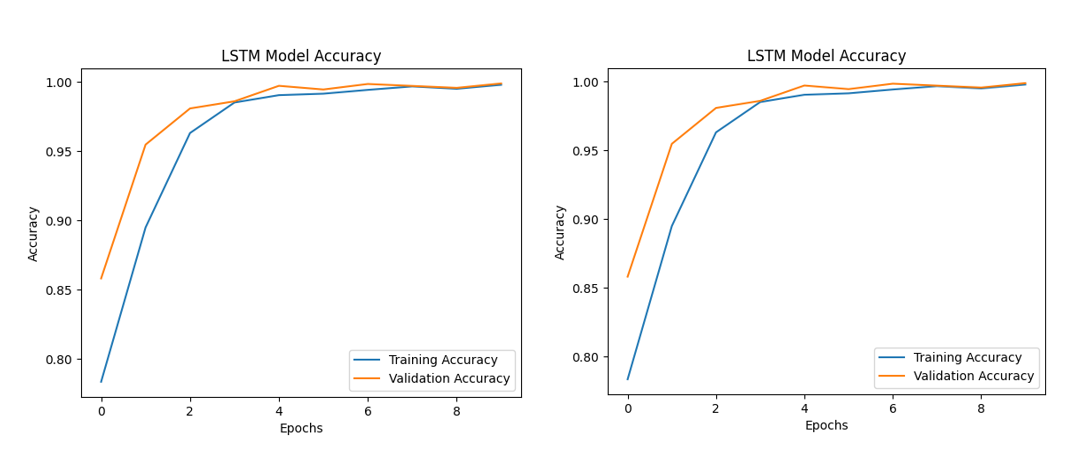}
\caption{Validation loss and Accuracy curve of LSTM} 
\label{fig:val}
\end{figure*}

\begin{figure*}[h!]
\centering
\includegraphics[width=1.5\columnwidth, height=0.3\textheight]{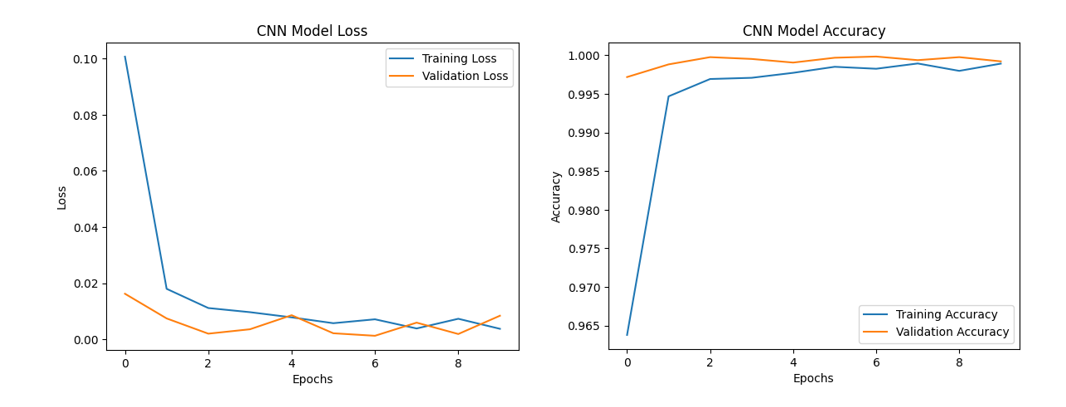}
\caption{Validation loss and Accuracy curve of CNN} 
\label{fig:val2}
\end{figure*}

It is evident from Figures \ref{fig:val}, \ref{fig:val1}, and \ref{fig:val2} that there is little variation in the validation accuracy and loss. Because of their stability, we may assume that these models have internalized the underlying pattern found in the data without being overfitted or underfitted. Since they show that the models can stay on course without nosediving, the little deviations shown here are encouraging for the models' predictive accuracy. Ultimately, these findings demonstrate the models' resilience in addressing the given job.

The scores on the table \ref{tab:metrics} for the deep learning models, CNN, LSTM and DNN are MCC (Matthews Correlation Coefficient) and Kappa. It got an MCC score of 99.46\% and a Kappa score of 99.46\%. The Kappa score of the LSTM model achieved value 99.76\% while it achieved an MCC score of 99.76\%. With an MCC score of 99.99\%, and Kappa of 100\%, the DNN model performed well. The values of these scores suggest that these models are making reliable predictions and that the DNN model performs the best on MCC and Kappa values, which indicates that the predicted values agree very well with actual values.

\begin{table}[h!]
    \centering
    \caption{MCC and Kappa Score of Deep Learning Models}
    \label{tab:metrics}
    \renewcommand{\arraystretch}{1.2} % Adjust row height for readability
    \setlength{\tabcolsep}{4pt} % Adjust column spacing
    \small % Reduce font size for compact display

    \begin{tabular}{|p{3cm}|p{2.5cm}|p{2.5cm}|} % Define column widths in cm
        \hline
        \textbf{Model} & \textbf{MCC Score (\%)} & \textbf{Kappa Score (\%)} \\ 
        \hline
        CNN  & 99.46 & 99.46 \\ 
        \hline
        LSTM & 99.76 & 99.76 \\ 
        \hline
        DNN  & 99.99 & 100.00 \\ 
        \hline
    \end{tabular}
\end{table}

\begin{figure}[h!]
\centering
\includegraphics[width=0.8 \columnwidth, height=0.3\textheight]{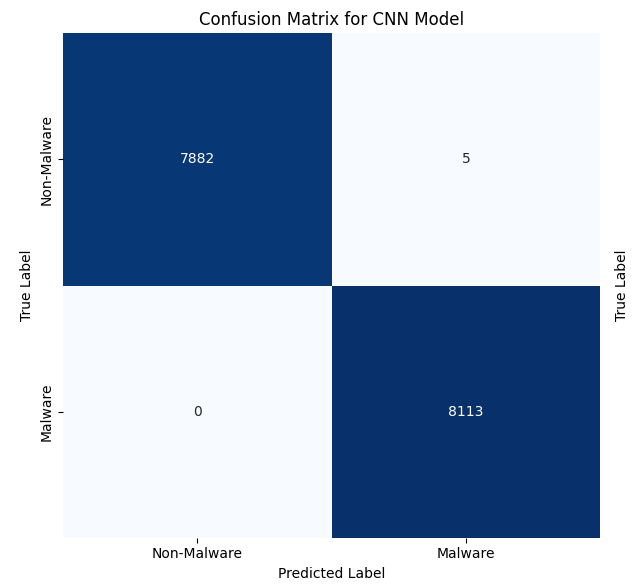}
\caption{Confusion Matrix of CNN} 
\label{fig:cc}
\end{figure}

\begin{figure}[h!]
\centering
\includegraphics[width=0.8 \columnwidth, height=0.3\textheight]{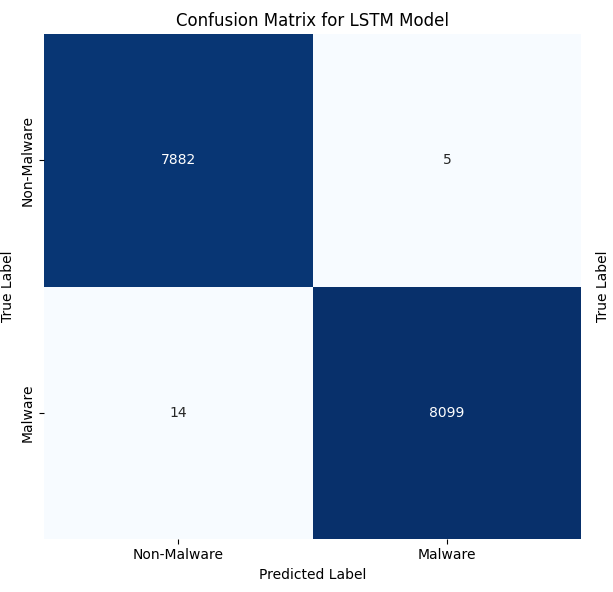}
\caption{Confusion Matrix of LSTM} 
\label{fig:cc2}
\end{figure}

\begin{figure}[H]
\centering
\includegraphics[width=0.8 \columnwidth, height=0.3\textheight]{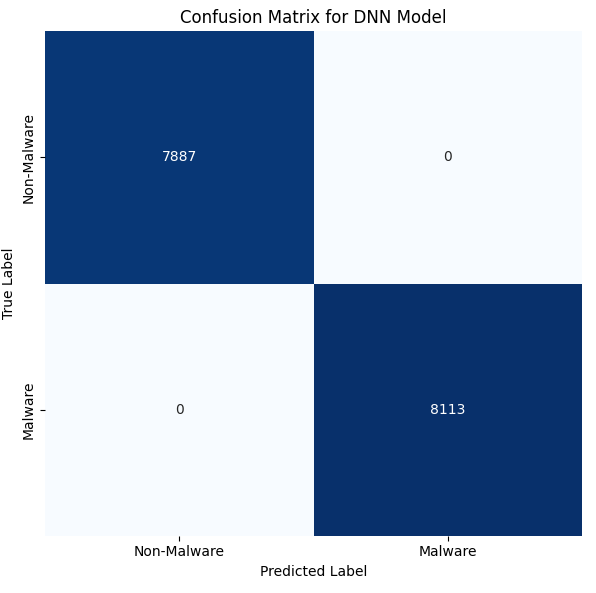}
\caption{Confusion Matrix of DNN} 
\label{fig:cc1}
\end{figure}

The DNN model also has a great capacity to detect all types of malware without any omissions, as seen in Figures \ref{fig:cc}, \ref{fig:cc2} and \ref{fig:cc1}, which shows the greatest accuracy in this evaluation. However, both CNN and LSTM have major drawbacks, including erroneous negative findings and, in the case of LSTM, false positive results. These findings show that in order to decrease misidentification and increase overall detection effectiveness, these models must be further refined and implemented.

\begin{figure*}[H]
\centering
\includegraphics[width=1.5\columnwidth, height=0.2\textheight]{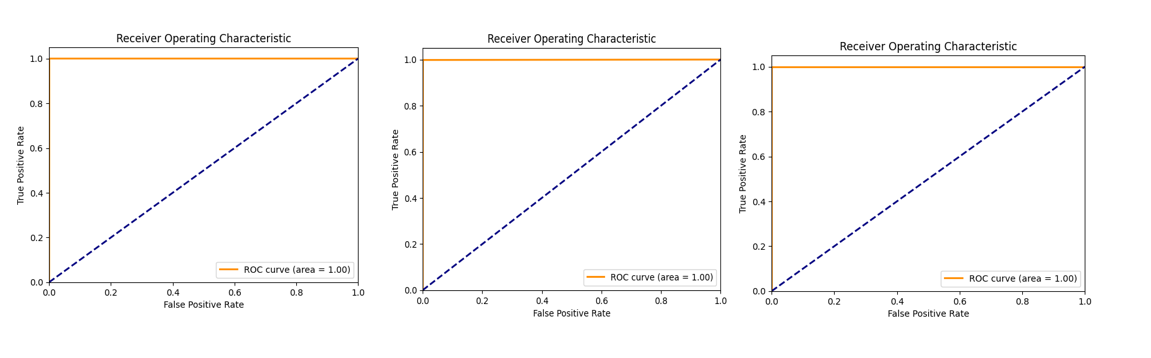}
\caption{ROC Curve of All Deep Learning Models}
\label{fig:rr}
\end{figure*}

Figure~\ref{fig:rr} shown that all the deep learning model have a perfect roc curve that all curves positioned in left top corner where the roc curve is flat. This just means that all the models perfectly classify malware and not malware.

\section{Proposed Approach}
The DNN model achieves top performance through 99.90\% accuracy together with 0.9993 AUC and 99.99\% MCC and 100\% Kappa scores which indicate highly reliable predictions. The malware detection system proves to be the most suitable because it achieves stable results without producing any errors. Due to its superior performance this proposed model stands as the top selection. The hyperparameter used in DNN is given in the following table \ref{tab:t}.

\begin{table}[ht]
\centering
\caption{DNN Model Hyperparameters}
\label{tab:t}
\renewcommand{\arraystretch}{1.2} % Increase row height
\setlength{\tabcolsep}{6pt} % Increase column spacing
\small % Keep text size small but readable

\begin{tabular}{|p{3.8cm}|p{3.8cm}|} % Increase column width for stretching
\hline
\textbf{Hyperparameter} & \textbf{Value} \\ \hline
Model Type & DNN \\ \hline
Optimizer & Adam \\ \hline
Loss Function & Binary Cross-Entropy \\ \hline
Activation (Hidden) & ReLU \\ \hline
Activation (Output) & Sigmoid \\ \hline
Epochs & 10 \\ \hline
Batch Size & 32 \\ \hline
Validation Split & 0.2 \\ \hline
Dropout Rate & 0.5 \\ \hline
Hidden Layer 1 Units & 128 \\ \hline
Hidden Layer 2 Units & 64 \\ \hline
Input Shape & Based on $X_{train}.shape[1]$ \\ \hline
\end{tabular}
\end{table}

\section{Comparative Analysis}

This current study surpasses most of the existing literature in terms of accuracy, achieving 99.99\% with our proposed DNN model summarized in Table \ref{tab:comparison_models}.

\begin{table}[ht]
\centering
\caption{Comparison with previous studies}
\resizebox{\columnwidth}{!}{
    \begin{tabular}{|l|l|c|}
        \hline
        \textbf{Study} & \textbf{Model Used} & \textbf{Best Accuracy(\%)} \\ \hline
        \cite{sym14112304} & Decision Tree & 99.00 \\ \hline
        \cite{10.1007/978-981-16-7618-5_53} & Random Forest & 99.44 \\ \hline
        \cite{DIVAKARLA2022148} & Deep Learning (EMBER dataset) & 87.76 \\ \hline
        \cite{informatics8010010} & Random Forest & 99.24 \\ \hline
        \cite{article} & LSVC & 99.743\\ \hline
        \cite{bawazeer2021malware} & CNN & 98.68 \\ \hline
        \textbf{This Study} & DNN & 99.99 \\ \hline
    \end{tabular}
}
\label{tab:comparison_models}
\end{table}

\section{Conclusion}
In this study we discuss how to invoke ML and DL techniques for malware detection with the aim of establishing comprehensive analysis of optimized approaches. As the level of malwares sophistication increases, traditional methods of detection are shown to be ineffective and need advanced technique. The results illustrate that the DNN is capable of identifying complex, developing threats with greater accuracy than any other model, 99.99\% and a perfect AUC of 100\%. Feature selection and preprocessing was highly effective for improving the model performance, which made the DNN the most reliable approach.

In the future, detecting what we call objects based on the combination of strengths of both ML and DL, and that has nothing to do with labels and classes, may improve detection accuracy even more. The performance can be further enhanced through transfer learning especially when there are only a few labeled data available. Likewise, an ability to develop real time malware detection systems, that adapt in real time to new threats through an API, is also critical. In order to develop more resilient cybersecurity, it is necessary to expand research to encompass various, rather than some datasets, and thus obtain a more complete picture of model performance.

\begin{flushleft}
\sloppy

\end{flushleft}

\end{document}